\documentclass{PoS}

\title{First results on QCD thermodynamics\\ with HISQ action}

\ShortTitle{First results on QCD thermodynamics with HISQ action}

\author{\speaker{Alexei Bazavov}$^a$
         and Peter Petreczky$^b$\thanks{This work has been supported in part by contracts DE-AC02-98CH10886
          and DE-FC02-06ER-41439           
          with the U.S. Department of Energy
          and contract 0555397 with the National Science
          Foundation. The numerical calculations have been performed
          using the USQCD resources at Fermilab as well as the BlueGene/L
          at the New York Center for Computational Sciences (NYCCS).
          We thank Z.~Fodor and S.~Katz for sending us the stout data.}
         { }[The HotQCD collaboration]
         \footnote{HotQCD Collaboration members are: A.~Bazavov, T.~Bhattacharya, 
M.~Cheng, N.H.~Christ, C.~DeTar, S.~Ejiri, S.~Gottlieb, R.~Gupta, 
U.M.~Heller, K.~Huebner, C.~Jung, F.~Karsch,
E.~Laermann, L.~Levkova, C.~Miao, R.D.~Mawhinney, P.~Petreczky, 
C.~Schmidt, R.A.~Soltz, W.~Soeldner, R.~Sugar, D.~Toussaint and P.~Vranas}\\
\llap{$^a$}Department of Physics, University of Arizona,
        Tucson, AZ 85721, USA\\
\llap{$^b$}Physics Department, Brookhaven National Laboratory, Upton, NY 11973, USA\\
}

\abstract{

We report on investigations of the chiral and deconfinement aspects
of the finite temperature transition in 2+1 flavor QCD using the
Highly Improved Staggered Quark (HISQ) action on lattices with
temporal extent $N_\tau=6$ and $N_\tau=8$. We have calculated several
physical observables, including the renormalized Polyakov loop, 
strangeness fluctuations, the renormalized 
chiral condensate and the chiral susceptibility in the crossover
region for physical values of the strange quark mass $m_s$ and light quark
masses $m_l=0.2m_s$ and $0.05m_s$. We compare our findings with previous
calculations that use different 
improved staggered fermion formulations: asqtad, p4
and stout.
}

\FullConference{The XXVII International Symposium on Lattice Field Theory\\
                 July 26-31, 2009\\
                 Peking University, Beijing, China}

\begin{document}

\section{Introduction}

Improved staggered fermion formulations are widely used to study 
QCD at non-zero temperatures and densities, see e.g. Ref. \cite{carleton,petr}
for recent reviews. The main reason for this
is the fact that they preserve a part of the chiral symmetry
of the continuum QCD. This makes the numerical simulations 
relatively inexpensive because due to absence of an additive mass
renormalization the Dirac operator is bounded from below, and also allows one
to study the chiral aspects of the finite temperature transition.
However, there are at least two problems with staggered fermion formulation.
The first one is the validity of the rooting procedure, i.e. the way to reduce
the number of tastes from four to one and the other is 
breaking of the taste symmetry
at finite lattice spacing. For the discussion of the validity of rooted 
staggered fermions see Refs. \cite{sharpe06,creutz07}. 
To reduce the taste violations smeared links are used in the staggered
Dirac operator and different staggered formulations, like
p4, asqtad and stout differ in the choice
of the smeared gauge links. The smeared gauge links in the p4 and asqtad actions
are linear combinations of single links and different staples 
\cite{karsch01,orginos} and therefore are not elements of SU(3) group.
It is known that projecting the smeared gauge fields onto the
SU(3) group greatly improves the taste symmetry \cite{anna}. The
stout \cite{fodor05} action and the HISQ action considered here implement the
projection of the gauge field onto SU(3) (or simply U(3)) group 
and thus achieve better
taste symmetry at a given lattice spacing. For studying QCD at high
temperature it is important to use discretization schemes which 
improve the quark dispersion relation, thus eliminating the 
tree level ${\cal O}(a^2)$ lattice artifacts in thermodynamic quantities.
The p4 and asqtad actions implement this improvement by introducing
3-link terms in the staggered Dirac operator.
In this contribution we
report on exploratory studies of QCD thermodynamics with the
HISQ action which removes tree level ${\cal O}(a^2)$ lattice artifacts
as well as has projected smeared links that greatly improve the taste symmetry.
We also compare our results with previous ones obtained
with the asqtad, p4 and stout actions \cite{milc04,eos005,pp005,fodor06,fodor09}.

\section{Action and run parameters}

The Highly Improved Staggered Quark (HISQ) action developed 
by the HPQCD/UKQCD collaboration \cite{Follana:2006rc}
reduces taste symmetry breaking and decreases the splitting
between different pion tastes by a factor of about three 
compared to the
asqtad action. The net result, as recent studies show
\cite{Bazavov:2009wm}, is that a HISQ ensemble at lattice
spacing $a$ has scaling violations comparable to ones 
in an asqtad ensemble at lattice spacing $2/3a$.

In this exploratory study we used the HISQ action 
in the fermion sector
and tree-level Symanzik improved gauge action without
the tadpole improvement. The strange quark mass $m_s$ 
was set to its physical value setting
the quantity $\sqrt{2m_K^2-m_\pi^2}=m_{\eta_{s\bar s}}\simeq \sqrt{2B m_s}$ 
to the physical  value 686.57~MeV. Two sets of ensembles have
been generated along the two lines of constant physics
(LCP): $m_l=0.2m_s$ and $m_l=0.05m_s$. In the first
set runs were performed on $16^3\times 32$
lattices at zero temperature and $16^3\times 6$ at finite
temperature, and in the second set on 
$32^4$ and $32^3\times 8$ lattices, correspondingly.
The parameters and statistics of these runs are
summarized in Table~\ref{tab_runs}. The molecular dynamics (MD)
trajectories have length of 1 time unit (TU) and
the measurements were performed every 5 TUs at zero
and 10 TUs at finite temperature. Normally, 
300 TUs were discarded for equilibration. The lattice
spacing has been determined by measuring the static
quark-anti-quark potential and using the Sommer scale
$r_0=0.469$ fm. The masses of several hadrons 
measured on zero-temperature ensembles fall into ranges
summarized in Table~\ref{tab_hadrons}.
\begin{table}
\centering
\begin{tabular}{|l|l|l|l|l||l|l|l|l|l|}
\hline
\multicolumn{5}{|c||}{
$N_\tau=6$, $0.2m_s$ LCP runs
} & 
\multicolumn{5}{|c|}{
$N_\tau=8$, $0.05m_s$ LCP runs
} \\ \hline
$\beta$ & $a$, fm & $am_s$ & TU, 0       & TU, $T$ &
$\beta$ & $a$, fm & $am_s$ & TU, 0       & TU, $T$ \\\hline
6.000   & 0.2297  & 0.115  & 3,000       & 6,000 &
6.354   & 0.1578  & 0.0728 & 1,095       & 2,590 \\
6.038   & 0.2212  & 0.108  & 3,000       & 6,000 &
6.423   & 0.1487  & 0.0670 & 1,295       & 2,490 \\
6.100   & 0.2082  & 0.100  & 3,000       & 6,000 &
6.488   & 0.1403  & 0.0620 & 1,495       & 2,690 \\
6.167   & 0.1954  & 0.091  & 3,000       & 6,000 &
6.550   & 0.1326  & 0.0582 & 1,395       & 3,380 \\
6.200   & 0.1895  & 0.087  & 3,000       & 6,000 &
6.608   & 0.1253  & 0.0542 & 1,460       & 3,100 \\
6.227   & 0.1848  & 0.084  & 3,000       & 6,000 &
6.664   & 0.1186  & 0.0514 & 1,095       & 5,000 \\
6.256   & 0.1800  & 0.081  & 3,000       & 6,000 &
6.800   & 0.1047  & 0.0448 & --          & 2,700 \\
6.285   & 0.1752  & 0.079  & 3,000       & 6,000 &
6.950   & 0.0921  & 0.0386 & --          & 1.400 \\
6.313   & 0.1708  & 0.076  & 3,000       & 6,000 &
7.150   & 0.0770  & 0.0320 & 1,095       & 1,190 \\
6.341   & 0.1665  & 0.074  & 3,000       & 6,000 &
        &         &        &             &       \\
6.369   & 0.1622  & 0.072  & 3,000       & 6,000 &
        &         &        &             &       \\
6.396   & 0.1582  & 0.070  & 3,000       & 6,000 &
        &         &        &             &       \\
6.450   & 0.1505  & 0.068  & 3,000       & 6,000 &
        &         &        &             &       \\
\hline
\end{tabular}
\label{tab_runs}
\caption{The parameters of the numerical simulations:
gauge coupling, strange quark mass and the number of time
units (TU), i.e. the number of MD trajectories for each run. Here
TU, 0 stands for the number of time units for zero-temperature runs,
while TU, $T$ is the number of time units for finite-temperature runs.}
\end{table}
\begin{table}
\centering
\begin{tabular}{|l|r|r|}
\hline
          & $0.2m_s$ LCP & $0.05m_s$ LCP \\\hline
$m_\pi$   & 306-312      & 156-161 \\
$m_K$     & 522-532      & 493-501 \\
$m_\rho$  & 850-883      & 760-808 \\
$m_N$     & 1130-1183    & 1014-1083\\
\hline
\end{tabular}
\label{tab_hadrons}
\caption{Ranges of masses (in MeV) of several hadrons
for the two sets of ensembles.}
\end{table}

\begin{figure}
\includegraphics[width=0.49\textwidth]{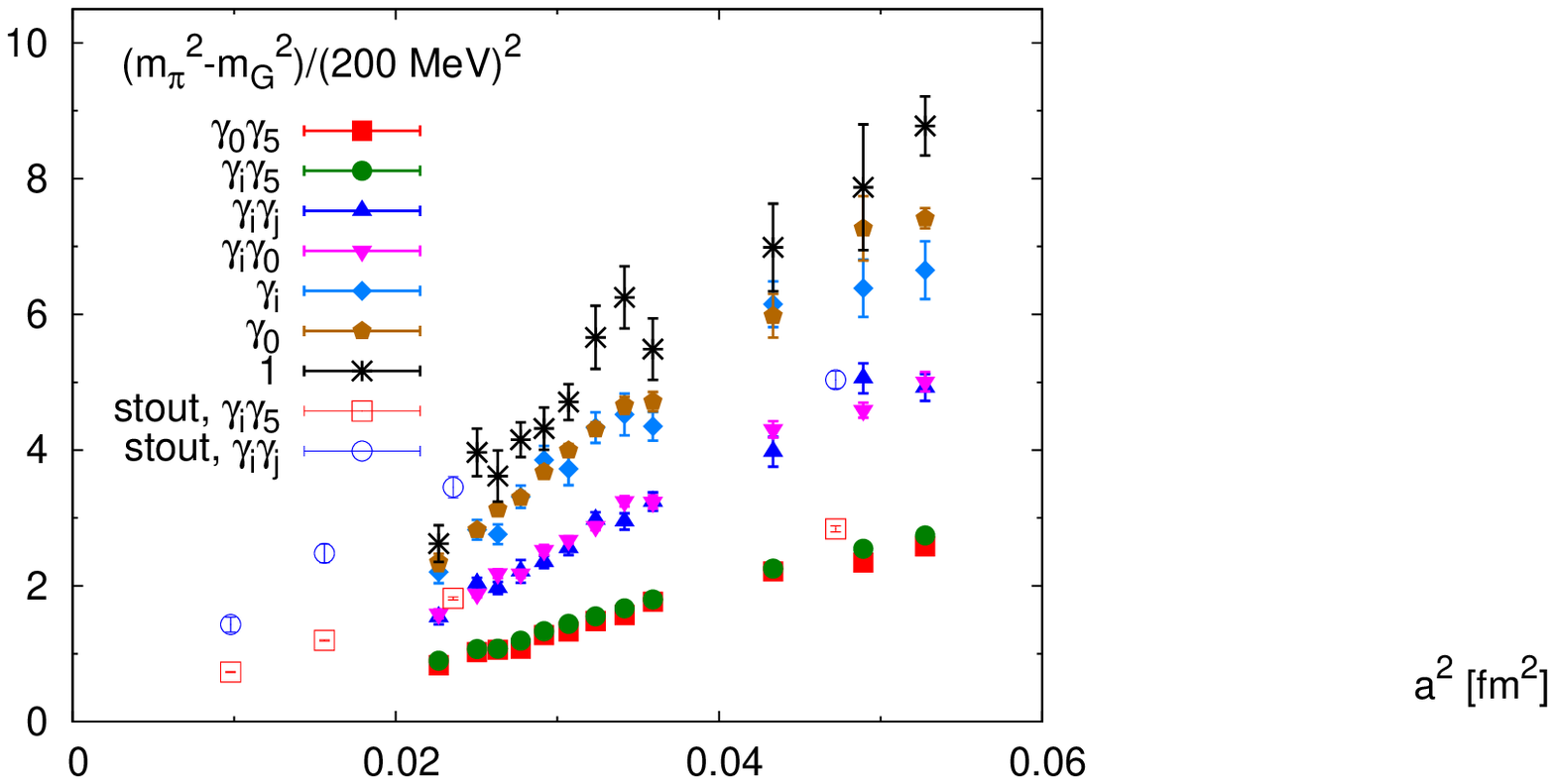}\hfill
\includegraphics[width=0.49\textwidth]{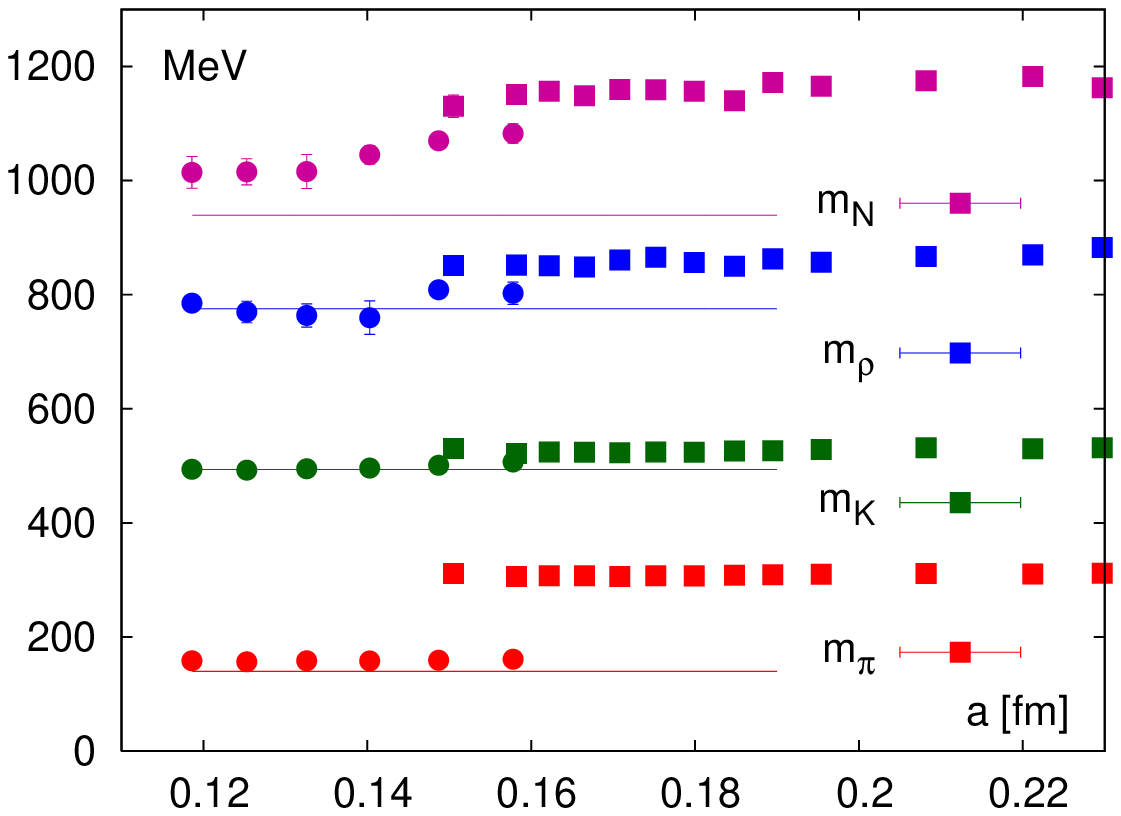}
\caption{The splitting between different pion multiplets (left) and
hadron masses (right) calculated for the HISQ action. In the right panel
squares indicate ensembles along the $0.2m_s$ LCP, and circles along
the $0.05m_s$ LCP.}
\label{fig:mass}
\end{figure}

As mentioned above the lattice artifacts for the HISQ action are significantly reduced
compared to the asqtad action. The taste violations are strongest in the pseudo-scalar
meson sector. Therefore we studied the splitting of pseudo-scalar meson masses in the
8 different multiplets. The results are shown in Fig. \ref{fig:mass} and compared to
the stout results. The splittings are 2 to 3 times smaller than in the calculations with
the asqtad action and also somewhat smaller than for the stout action.
The smaller taste violations also result in better scaling of other hadron masses.
In Fig.~\ref{fig:mass} we also show the kaon, rho-meson and nucleon masses. As one 
can see for the small, almost physical, value of the light quark mass we get reasonable
agreement between the lattice results and the experimental values.

\section{Deconfinement transition}
\begin{figure}
\centering
\includegraphics[width=0.5\textwidth]{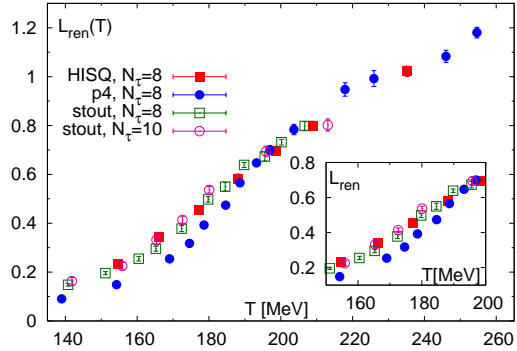}
\caption{The renormalized Polyakov loop as function of the temperature calculated
for the HISQ, stout and p4 actions at the physical value of light quark masses.}
\label{fig:poly}
\end{figure}

\begin{figure}
\includegraphics[width=0.49\textwidth]{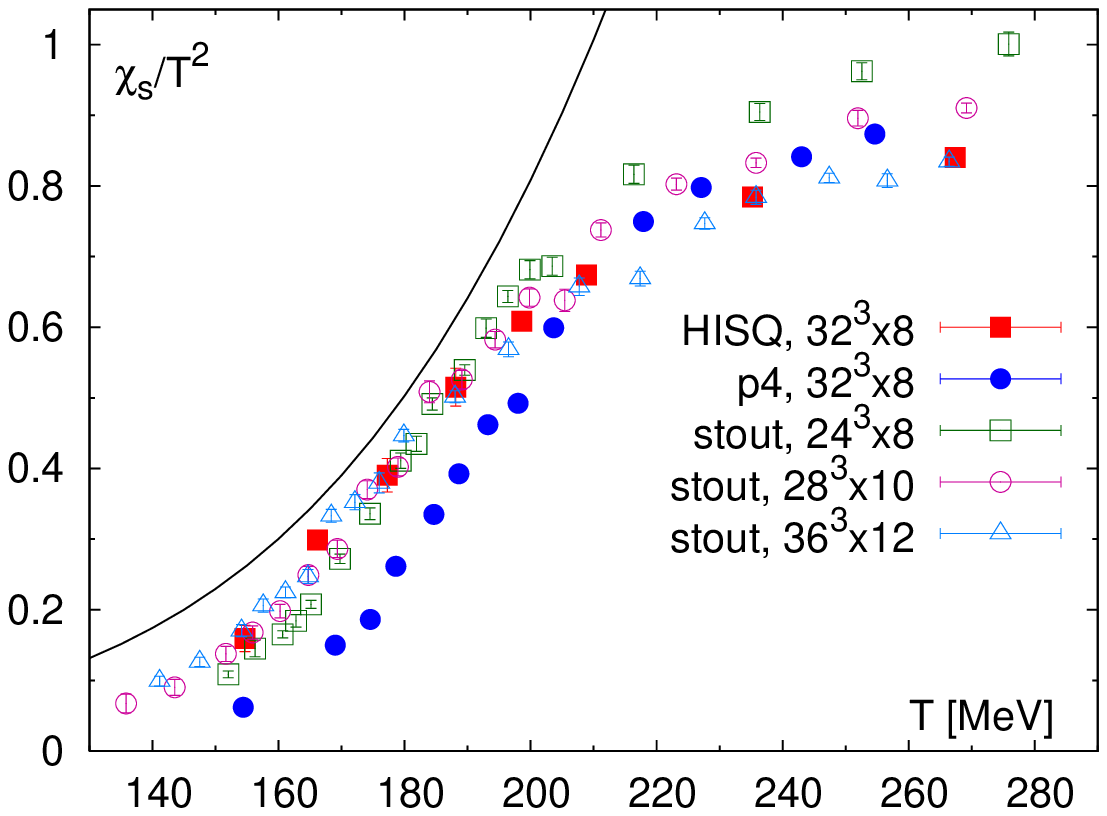}
\hfill
\includegraphics[width=0.49\textwidth]{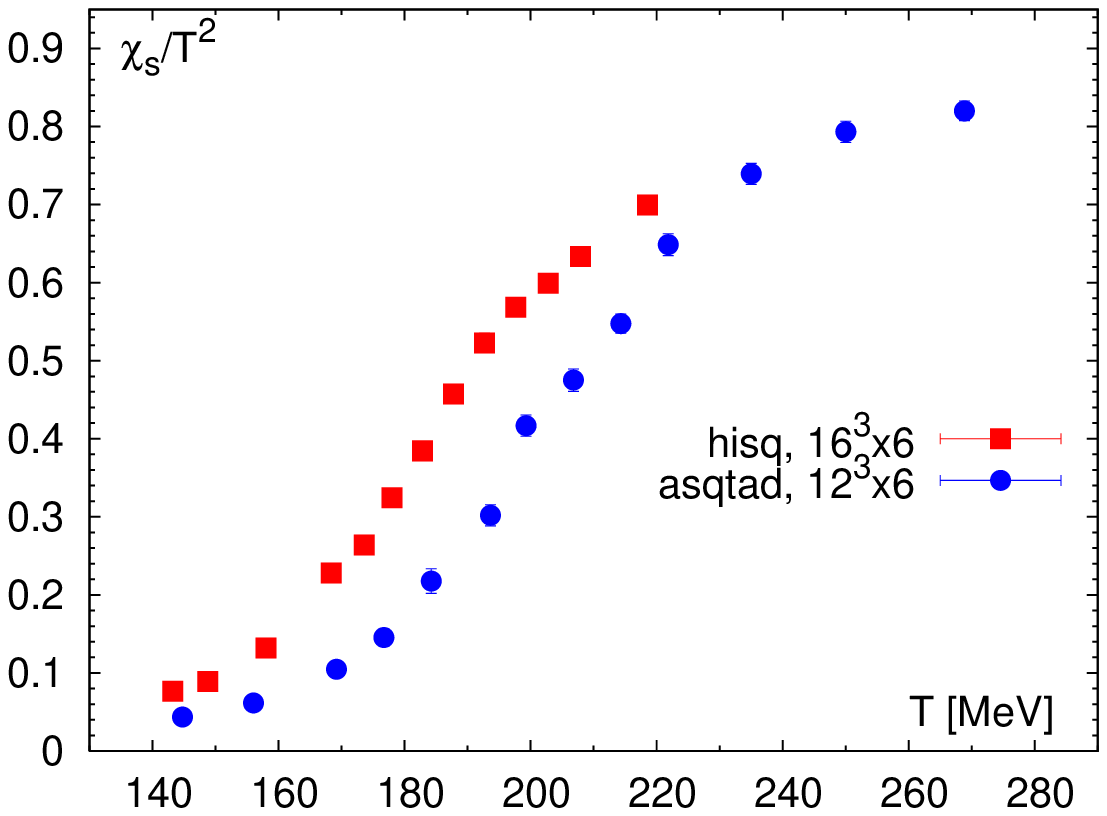}
\caption{The strangeness fluctuations $\chi_s(T)$ calculated for $m_q=0.05m_s$ (left) and
$m_q=0.2m_s$ (right). For the physical quark mass $m_q=0.05m_s$ we compare our results with
previous calculations performed with the stout and p4 actions.
The solid line in the left plot shows the prediction of the HRG model.
}
\label{fig:chis}
\end{figure}

The deconfinement transition is usually studied using the renormalized Polyakov loop
\cite{fodor06,okacz02,p4eos,hotqcd}. 
It is related to the free energy of a static quark anti-quark pair at infinite separation
$F_{\infty}(T)$
\begin{equation}
L_{ren}(T)=\exp(-F_{\infty}(T)/(2 T)),
\end{equation}
obtained from the bare Polyakov loop as
\begin{eqnarray}
&
\displaystyle
L_{ren}(T)=z(\beta)^{N_{\tau}} L_{bare}(\beta)=
z(\beta)^{N_{\tau}} \left<\frac{1}{3}  {\rm Tr } 
\prod_{x_0=0}^{N_{\tau}-1} U_0(x_0,\vec{x})\right >.
\end{eqnarray}
Here the multiplicative renormalization constant $z(\beta)$ is related to the 
additive normalization of the potential $c(\beta)$ as $z(\beta)=\exp(-c(\beta)/2)$.
To make the comparison with the stout results easy 
here we chose the normalization constant $c(\beta)$
such that the potential is zero at distance $r=r_0$. This choice is different from
the one used in Refs. \cite{p4eos,hotqcd}. Our results for the quark mass $m_q=0.05m_s$
on $32^3 \times 8$ lattices 
are shown in Fig. \ref{fig:poly} and compared to the stout and p4 calculations.
As one can see the HISQ calculations agree reasonably well 
with the stout results if the scale is set by $r_0$ in the stout calculations
\footnote{In what follows we will show the stout results using the temperature scale
set by $r_0$ instead of the kaon decay constant $f_K$. We used the published
values of $r_0$ and $f_K$ in Refs. \cite{fodor06,fodor09} to convert the two scales.}.
On the other hand, the renormalized Polyakov loop calculated with the p4 action
is considerably smaller at low temperatures. At temperatures $T>200$ MeV we see 
good agreement for different actions.
The fluctuation of strangeness is also a good indicator for the deconfinement transition. It can be defined
as the second derivative of the free energy density with respect to the strange quark
chemical potential
\begin{equation}
\frac{\chi_s(T)}{T^2}=
\left.\frac{1}{T^3 V}\frac{\partial^2 \ln Z(T,\mu_s)}{\partial (\mu_s/T)^2}
\right|_{\mu_s=0}.
\end{equation}
At low temperatures 
strangeness is carried by massive hadrons and therefore strangeness fluctuations are
suppressed. At high temperatures strangeness is carried by quarks and 
the effect of the non-zero strange quark mass is small. Therefore, in the transition region 
the strangeness
fluctuation rises and eventually reaches a value close to that of an ideal quark gas. 
Our numerical results for strangeness fluctuations are shown in Fig. \ref{fig:chis} for
two values of the light quark masses, $m_q=0.05m_s$ and $0.2m_s$. 
In the case of the heavier quark mass we compare our results
with the ones obtained with the asqtad action for $m_q=0.2m_s$. 
As one can see the strangeness fluctuations are larger
for the HISQ action at low temperatures, $T<210$ MeV. 
In other words, the transition region in the HISQ calculation shifts toward
smaller temperatures. This behavior is in fact expected. Due to smaller 
taste symmetry violation pseudo-scalar masses as
well as other hadron masses are smaller and therefore 
strangeness fluctuations are larger.
For the smaller (physical) quark mass we compare our calculations 
with the results obtained using the p4 action \cite{eos005,pp005}
and the stout action \cite{fodor06,fodor09}. At low temperatures, $T<200$ MeV 
the HISQ results are significantly larger than the p4
results but are in good agreement with the stout results. 
At high temperatures, $T>200$ MeV the strangeness fluctuations
calculated with the HISQ action are in reasonable agreement 
with the p4 results as well as the $N_{\tau}=12$ stout results. We also
see that in this temperature region the stout results show some cutoff 
($N_{\tau}$) dependence. This is due to the fact
that the tree level ${\cal O}(a^2)$ lattice artifacts are not removed in the stout action.
In the low temperature region we expect that the Hadron Resonance Gas (HRG) model 
gives a reasonably good
description of the thermodynamic quantities, including strangeness fluctuations. 
Therefore in Fig. \ref{fig:chis}
we show the prediction of the HRG model. As one can see the lattice results fall below the HRG model results.
The discrepancy between the HRG model and the lattice data becomes larger at smaller temperatures, although the model
is expected to be more reliable at smaller temperatures.
This is presumably due to taste violation, especially in the pseudo-scalar meson sector. Due to non-negligible splitting
in the pseudo-scalar meson mass the contribution of  kaons to strangeness fluctuations appears to be smaller than expected
in the continuum. 

\section{The chiral transition}

\begin{figure}
\includegraphics[width=0.49\textwidth]{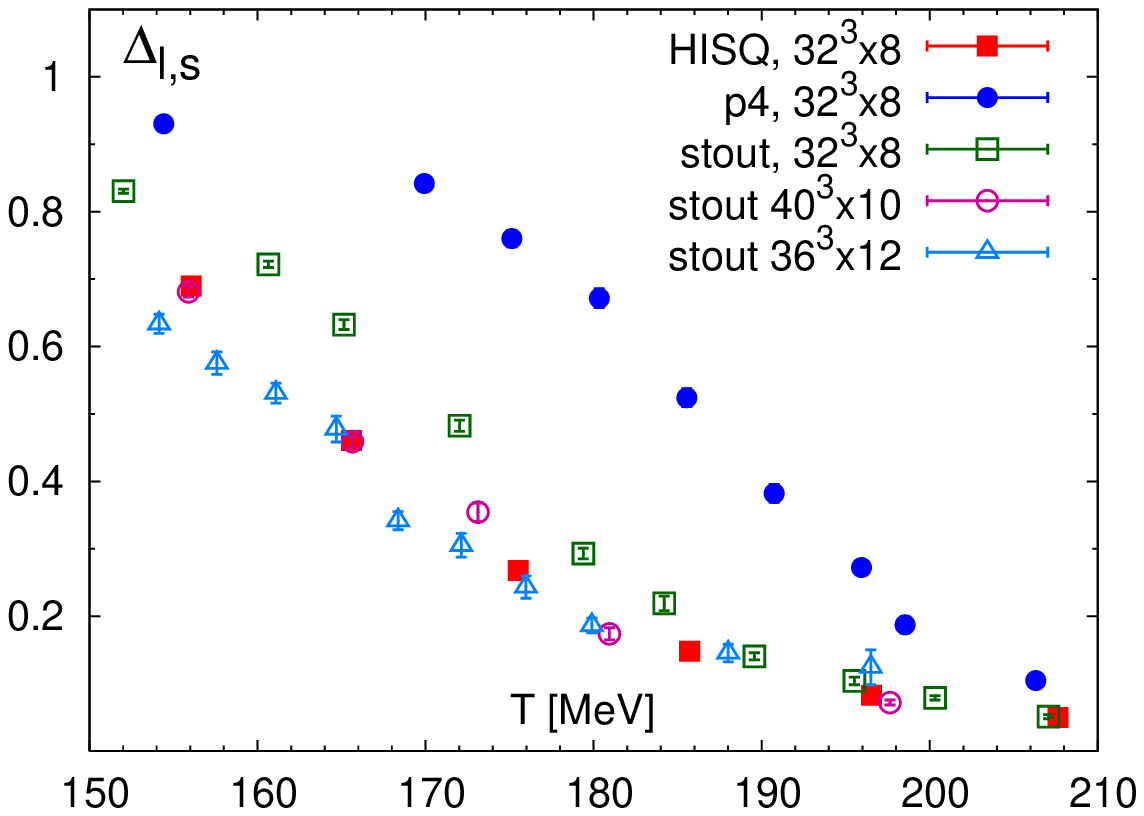}\hfill
\includegraphics[width=0.49\textwidth]{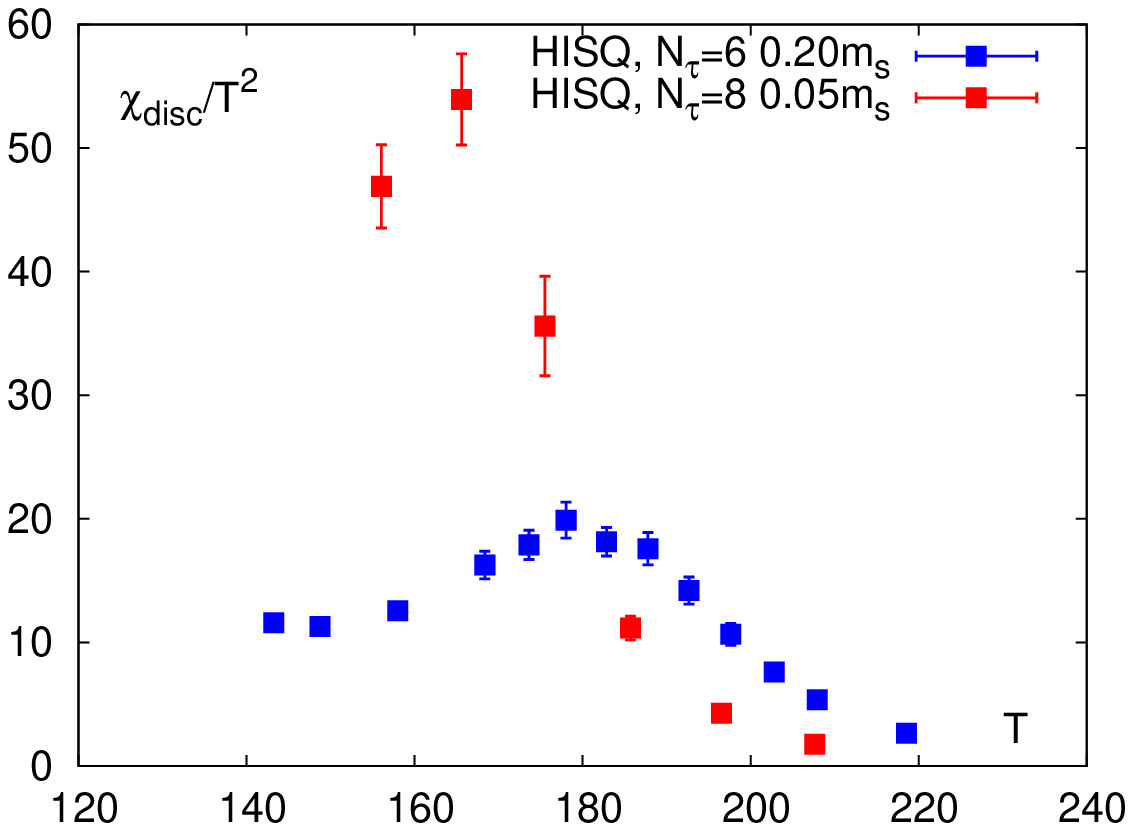}
\caption{The subtracted chiral condensate (left) and the disconnected part of the chiral
susceptibility (right). For the subtracted chiral condensate we also compare our results
with calculations performed for stout and p4 actions. Note that in the right panel 
we show the disconnected chiral susceptibility for the HISQ action for two LCPs:
$0.2m_s$ and $0.05m_s$.}
\label{fig:pbp}
\end{figure}

In the limit of zero light quark masses QCD has a chiral symmetry and
the finite temperature transition is a true phase transition. The order
parameter for this transition is the light chiral condensate 
$\langle \psi \bar \psi \rangle_{l}$.
However, even at finite values of the quark mass the chiral condensate 
will show a rapid change in the
transition region indicating an effective restoration of the chiral symmetry. 
Since the chiral condensate
has an additive ultraviolet renormalization we consider 
the subtracted chiral condensate \cite{p4eos,hotqcd}
\begin{equation}
\Delta_{l,s}(T)=\frac{\langle \bar\psi \psi \rangle_{l,\tau}-\frac{m_l}{m_s} \langle \bar \psi \psi \rangle_{s,\tau}}
{\langle \bar \psi \psi \rangle_{l,0}-\frac{m_l}{m_s} \langle \bar \psi \psi \rangle_{s,0}}.
\end{equation}
Here the subscripts $l$ and $s$ refer to light and strange quark condensates 
respectively, while the
subscripts $0$ and $\tau$ refer to the zero and finite temperature cases.
In Fig. \ref{fig:pbp} we show the renormalized chiral condensate calculated 
with the HISQ action
and compare it with results obtained with the stout action \cite{fodor09} 
as well as with the p4
results \cite{eos005,pp005}. Our results agree resonably well with the $N_{\tau}=8$
 stout results and the agreement
between the HISQ and stout results is even better if one considers 
$N_{\tau}=10$ and $N_{\tau}=12$. On the other hand,
the subtracted chiral condensate is considerably smaller than for the p4 action. This is presumably due
to the larger taste violating effects in the p4 case. We also considered the fluctuations of the chiral condensate, which
is just the disconnected part of the chiral susceptibility $\chi_{disc}$. The corresponding results are shown in Fig. \ref{fig:pbp}.
For true phase transition the chiral susceptibility should show a peak at the critical temperature. The numerical results
indeed show some peak-like structure. However, as discussed in Ref. \cite{gold} it is not easy to differentiate between
the increase in the chiral condensate due to nearby critical point and the effect of the Goldstone modes. 
Certainly, much more detailed studies are needed here. 

\section{Conclusions}

In this contribution we discussed first results on QCD thermodynamics with the HISQ action. The use
of this action allows one to minimize cutoff effects in thermodynamic quantities due to lattice
artifacts in the quark dispersion relation and taste symmerty breaking. In fact, the HISQ action
gives the smallest taste violation in the pseudo-scalar meson sector if measured though the mass splittings.
We have calculated the renormalized Polyakov loop, the chiral condensate and the strangeness fluctuations on
$N_{\tau}=6$ and $N_{\tau}=8$ lattices. 
We find good agreement between our results and the ones obtained with the stout action 
in the low temperature region,
$T<200$ MeV. At the same time present results disagree with the p4 results. 
In the high temperature region, where there are visible cutoff effects 
in the stout calculations the HISQ results agree
resonably well with the p4 results as well as with the stout results obtained on $N_{\tau}=12$.

\end{document}